\documentclass[twocolumn,showpacs,floatfix,superscriptaddress,amsmath,amssymb,prb]{revtex4}
\usepackage{mathrsfs}
\usepackage{txfonts}
\usepackage{amssymb}
\usepackage{graphicx}
\usepackage{hyperref}
\usepackage{ulem}
 \usepackage{overpic}
 \usepackage{psfrag}
\usepackage{tabularx}
\usepackage{array}
\usepackage{placeins}
\usepackage{color}

\makeatletter

\begin{document}
\title{The antiferromagnetic cross-coupled spin ladder:
quantum fidelity and tensor networks approach}

\author{Xi-Hao Chen}
\affiliation{Centre for Modern Physics, Chongqing University,
Chongqing 400044, The People's Republic of China}
\affiliation{Department of Physics, Chongqing University, Chongqing
400044, The People's Republic of China}

\author{Sam Young Cho}
\affiliation{Centre for Modern Physics, Chongqing University,
Chongqing 400044, The People's Republic of China}
\affiliation{Department of Physics, Chongqing University, Chongqing
400044, The People's Republic of China}

\author{Murray T. Batchelor}
\affiliation{Centre for Modern Physics,
Chongqing University, Chongqing 400044, The People's Republic of China}
\affiliation{Mathematical Sciences Institute and Department of Theoretical Physics,
Research School of Physics and Engineering, Australian National University, Canberra ACT 0200, Australia}

\author{Huan-Qiang Zhou}
\affiliation{Centre for Modern Physics, Chongqing University,
Chongqing 400044, The People's Republic of China}
\affiliation{Department of Physics, Chongqing University, Chongqing
400044, The People's Republic of China}

\begin{abstract}
We investigate the phase diagram of the cross-coupled Heisenberg spin ladder with antiferromagnetic couplings.
For this model there have been conflicting results for the existence of the columnar dimer phase, which was predicted
on the basis of weak coupling field theory renormalisation group arguments.
The numerical work on this model has been based on various approaches, including exact diagonalization,
series expansions and density-matrix renormalization group calculations.
Using the recently developed tensor network states and ground-state fidelity approach for quantum spin ladders we
find no evidence for the existence of the columnar dimer phase.
We also provide an argument based on the symmetry of the Hamiltonian which suggests that the phase diagram
for antiferromagnetic couplings consists of a single line separating the rung-singlet and Haldane phases.
\end{abstract}


\maketitle

\section{Introduction}

\begin{figure}[b]
\includegraphics[width=0.25\textwidth]{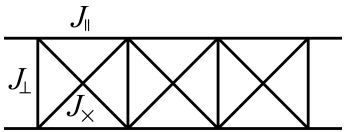}
\caption{
The cross-coupled two leg spin ladder.
}
\label{fig1}
\end{figure}

Geometric frustration induced by competing interactions in quantum systems produces
increased quantum fluctuations resulting in
a range of distinct quantum phases with potentially rich phase diagrams.\cite{frust}
Such systems can be low-dimensional if there are a sufficient number of
competing interactions, which need to be comparable in magnitude to induce
frustration.
The example which we examine here is the cross-coupled two-leg Heisenberg spin ladder.
Spin ladder systems are of considerable importance in condensed matter physics
and have thus been widely investigated,
with a number of known ladder compounds realised in experiments.\cite{Ladders}
The cross-coupled spin ladder has Hamiltonian
\begin{eqnarray}\label {ham}
H(J_\parallel,J_\perp,J_\times) &=& J_{\parallel} \sum_i (S_{1,i}\cdot{S_{1,i+1}} + S_{2,i}\cdot{S_{2,i+1}} ) \nonumber\\
&& + \, J_{\perp} \sum_i S_{1,i}\cdot{S_{2,i}} \nonumber\\
&& + \, J_{\times} \sum_{i}(S_{1,i} \cdot S_{2,i+1} + S_{2,i} \cdot S_{1,i+1}),
\end{eqnarray}
where $S_{1,i}$ and $S_{2,i}$ are spin-1/2 operators on rung $i$ and legs 1 and 2.
The parameters $J_{\parallel}$, $J_{\perp}$ and $J_{\times}$ are the intra leg,
inter leg (rung) and cross exchange couplings, respectively.
The spin ladder couplings are depicted in Fig.~\ref{fig1}.
We are most interested in competing antiferromagnetic couplings, for
which there have been conflicting results for the precise nature of the phase diagram.

The cross-coupled ladder model in Eq. (\ref{ham}) has been
investigated, using numerous methods, for over already two
decades.\cite{G91,BG93,X95,KO96,ZKO98,KM98,AEN00,KFSS00,HMT00,W00,FLS01,
HHR01,NT03,N03,SB04,HGCY06,AR08,KLS08,LWT08,HS10,LSSLDZ12,BLNS12,LDW12}
After some debate it now appears settled that the phase diagram for antiferromagnetic couplings
consists of a phase boundary separating rung-singlet and Haldane phases.
As emphasized,\cite{KLS08} the entire region above the transition line is in the rung-singlet phase,
being continuously related to the exactly solvable
line $J_\times = J_\parallel$ along which an exact rung-singlet state is the ground-state in this phase.
The entire region below the transition line is in the Haldane phase,
being continuously related to the point $(J_\times/J_\parallel = 1,
J_\perp/J_\parallel =0)$ at which the ladder is equivalent to a
spin-1 chain.\cite{G91,BG93,X95,KO96,ZKO98}

The debate centred over the existence or not of a columnar dimer (CD) phase
intermediate between the rung-singlet and Haldane phases which
was predicted on the basis of weak coupling field theory renormalisation group arguments.\cite{SB04}
This phase is between $J_\perp \simeq 2J_\times - (5J_\times^2)/(\pi^2 J_\parallel)$ and
$J_\perp \simeq 2J_\times - (J_\times^2)/(\pi^2 J_\parallel)$.\cite{SB04}
Subsequent numerical work motivated by this prediction disputed the existence of the CD phase.
\cite{HGCY06,KLS08}
On the other hand,
further numerical work provided evidence supporting the existence of the CD phase.\cite{LWT08,HS10}
In particular,
the CD phase was claimed to exist in the range $0.373 \le J_\perp \le 0.386$ for $J_\times=0.2$.\cite{HS10}
This point was further addressed by more extensive density-matrix renormalization group (DMRG) calculations in the
range $0.36 \le J_\perp \le 0.4$ for $J_\times=0.2$, from which it was concluded that the CD phase does not appear.\cite{BLNS12}
The most recent work highlighted the fundamental limitations of the DMRG method when applied to the cross-coupled ladder.\cite{BLNS12}
Nevertheless, the existence of the CD phase in this model is beyond dispute for ferromagnetic couplings,
as confirmed by a number of studies.\cite{HS10,LSSLDZ12,LDW12}

An efficient tensor network algorithm has been developed for spin ladder systems which
generates ground-state wave functions for infinite length quantum spin ladders.\cite{LSSLDZ12}
In general, tensor networks have proven to be a convenient means to represent wave functions of
quantum many-body lattice systems,\cite{TNreview}
and in particular, to calculate quantum fidelity as a measure of quantum state distinguishability.
 The ground-state fidelity per lattice site based on tensor networks
 has thus been established as a
 powerful tool to study quantum phase transitions regardless of what type of
 internal order exists in the quantum states.
 It has been demonstrated to directly
 (i) capture phase transition
 points due to changes of ground-state wave functions in the vicinity
 of quantum critical points \cite{ZB08,ZZL08,ZWLZ10,ZOV08}
 and (ii) detect degenerate ground-states due
 to spontaneous symmetry breaking. \cite{SHLC13,DCBZ14,WCB15}
The quantum fidelity approach can thus be utilized to provide evidence for the
existence or not of the CD phase, because by its definition,\cite{SB04,HS10,LSSLDZ12,LDW12} the
CD phase is to be induced by a spontaneous breaking of lattice translational symmetry.
 Motivated by the conflicting results for the existence of the CD phase in the antiferromagnetic regime
 and the advantages of the quantum fidelity and tensor networks approach,
  we have examined the cross-coupled ladder model,
 paying particular attention to the previously studied line $J_\times=0.2$.

The paper is organised as follows.
In Sec.~II we
outline the numerical procedure for the tensor network based quantum fidelity approach for the
cross-coupled spin ladder
and define the relevant order parameters.
Our numerical results using this method are also presented.
Concluding remarks are given in Sec.~III.

\section{Numerical procedure and results}
The ground-state fidelity and tensor network states approach has been developed
to study the criticality and phase diagrams of ladder models in Ref. \onlinecite{LSSLDZ12}.
We refer to that paper for technical details.
The tensor network representation suitable to describe the ground-state wave function
of an infinite-size spin ladder
is a particular case of projected entangled-pair states (PEPS)\cite{VC04}
adapted to the geometry of the spin ladder system.
For the two-leg ladder one needs to update four different four-index tensors.
In this prescription the updating procedure is closely connected to
the infinite PEPS algorithm\cite{JOVVC08} and the translationally
invariant MPS algorithm.\cite{PVV11}
The computational cost scales as $\mathbb{D}^6$ where $\mathbb{D}$ is the truncation dimension.
For the two-leg Heisenberg spin ladder it was demonstrated that the tensor network approach
could very accurately determine
the ground-state wave function, with truncation dimension $\mathbb{D}=6$
outperforming the DMRG results.\cite{LSSLDZ12}

\begin{figure}
\begin{center}
 \begin{overpic}[width=0.4\textwidth]{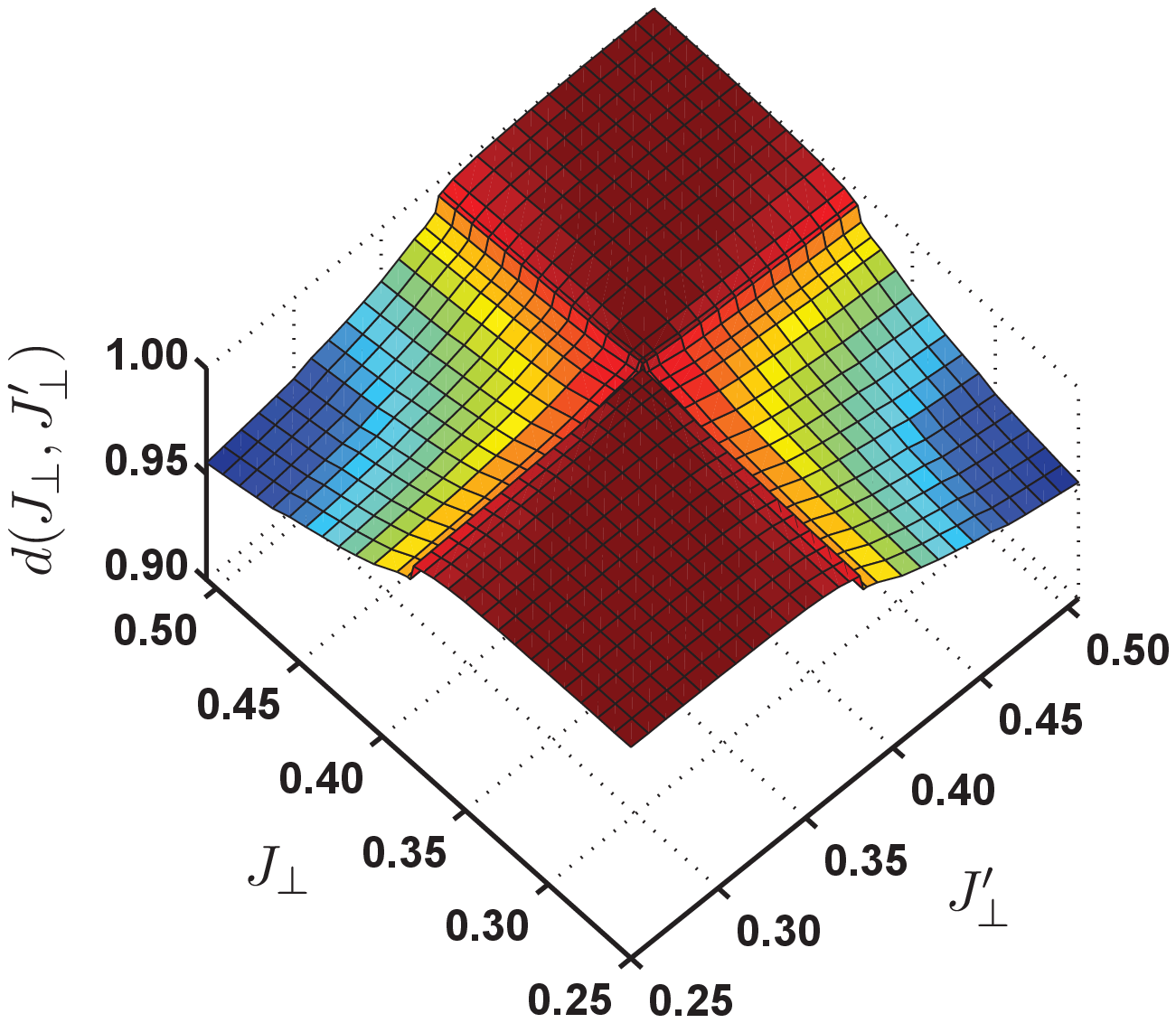}
 \put(0,80){(a)}
  \end{overpic}
 \begin{overpic}[width=0.4\textwidth]{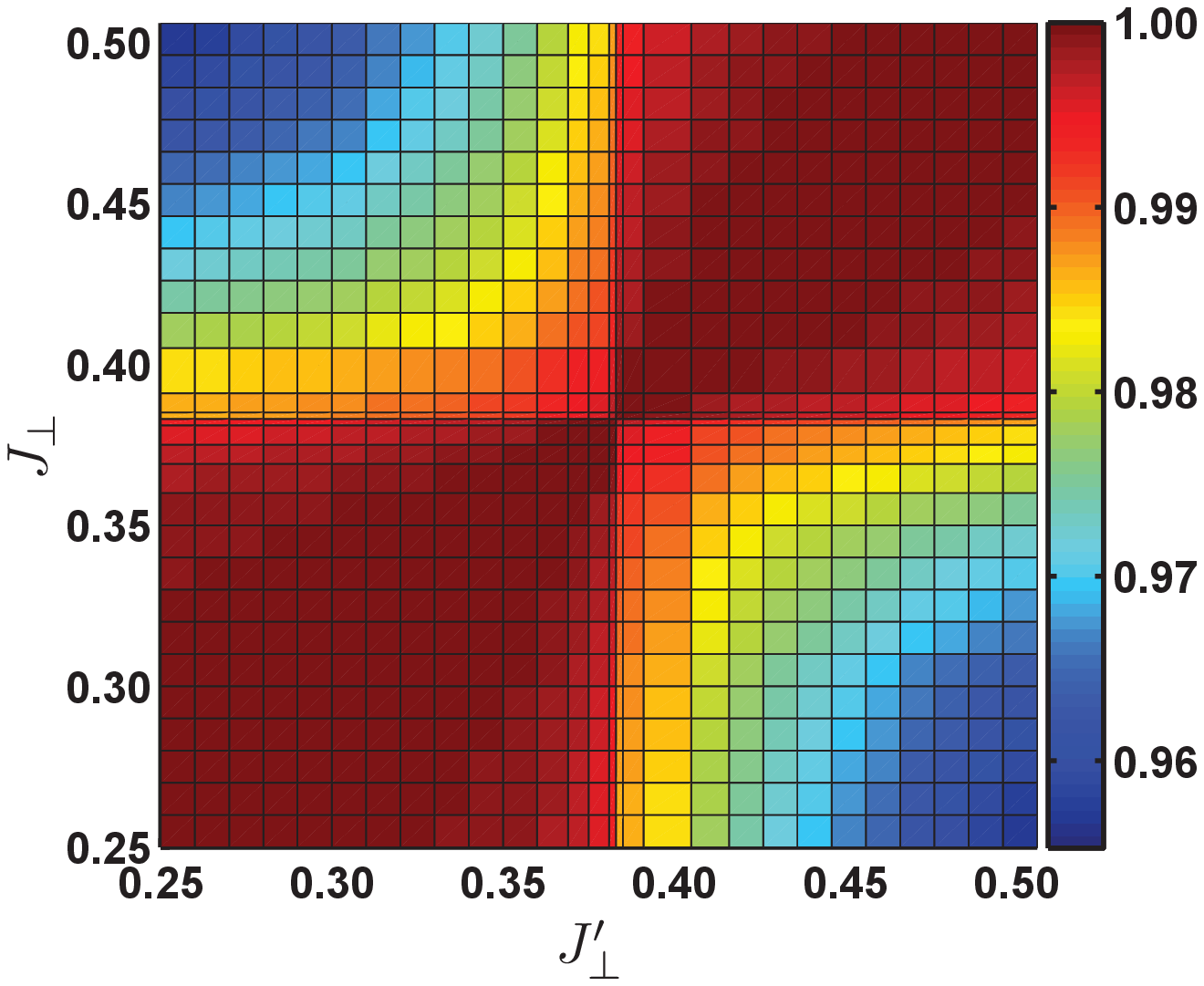}
 \put(0,80){(b)}
  \end{overpic}
 \end{center}
\caption{(color online) The ground-state fidelity per lattice site
$d(J_{\perp},J_{\perp}^{\prime})$ as a function of $J_{\perp}$ and
$J_{\perp}^{\prime}$ for the cross-coupled spin ladder with coupling
$J_\times = 0.2$. (a) A continuous phase transition point
$J_{\perp}^{c}\simeq0.383$ is identified as a pinch point
$(J_{\perp}^{c},J_{\perp}^{c})$ on the fidelity surface. Two
different phases are thus identified: the Haldane phase for
$J_{\perp}<J_{\perp}^{c}$ and the rung singlet phase for
$J_{\perp}>J_{\perp}^{c}$. (b) The corresponding planar view. }
\label{fig2}
\end{figure}

\subsection{Ground-state fidelity per lattice site}

\subsubsection{Phase transition point}

 For the cross-coupled spin ladder in Eq. (\ref{ham}) we set
 $J_\parallel=1$, fix $J_\times$ and vary $J_{\perp}$ as control
 parameter.
 With a randomly chosen initial state in the tensor network representation,
 the updating algorithm of the tensors until the saturation of the energy in
 Ref. \onlinecite{LSSLDZ12}
 allows us to obtain the ground-states of the spin ladder.
 For a given $J_\perp$ with $J_\times=0.2$,
 we calculate the ground-state $|\psi({J_{\perp}})\rangle$.
 In general,
 one can consider the quantum fidelity
 $F(|\psi({J_{\perp}})\rangle,|\phi\rangle)=|\langle\psi({J_{\perp}})|\phi\rangle|$
 which is the overlap function between the calculated ground-state $|\psi({J_{\perp}})\rangle$ and an arbitrary reference state
$|\phi\rangle$.
 The quantum fidelity scales as $F(|\psi(J_{\perp})\rangle,|\phi\rangle)
 \sim d(|\psi(J_{\perp})\rangle,|\phi\rangle)^L$,
 where $L$ is the system size.
 Following Ref.~\onlinecite{SHLC13}, the fidelity per lattice site can be defined as
\begin{equation}
 \ln d(\psi(J_{\perp})\rangle,|\phi\rangle)
 = \lim_{L \rightarrow \infty} \frac {\ln
  F(|\psi(J_{\perp})\rangle,|\phi\rangle)}{L}.
 \label{FLS}
\end{equation}
 It satisfies properties inherited from the characteristics of the
 fidelity $F(|\psi(J_{\perp})\rangle,|\phi\rangle)$, namely (i) normalization
 $d(|\psi(J_{\perp})\rangle=|\phi\rangle,|\phi\rangle)=1$
 and (ii) range $0\leq d(|\psi(J_{\perp})\rangle,|\phi\rangle)\leq{1}$.

 When the arbitrary reference is chosen as a ground-state of the spin ladder,
 the ground-state fidelity per lattice site has been demonstrated to
 be a universal marker for detecting the existence of quantum phase
 transitions, which are characterized by a pinch point on the
 fidelity surface,
 irrespective of the dimensionality of the
 system.\cite{ZB08,ZZL08,ZOV08,ZWLZ10}
 For our system, then,
 one can choose the reference state $|\phi\rangle = |\psi({J'_{\perp}})\rangle$
 to find a pinch point, i.e., a phase transition point.
 In this case, the fidelity per lattice site satisfies one more property, namely
(iii) symmetry
$d(J_{\perp},J_{\perp}^{\prime})=d(J_{\perp}^{\prime},J_{\perp})$.
 In Fig. \ref{fig2}, we plot the fidelity surface
 $d(|\psi({J_{\perp}})\rangle, |\psi({J'_{\perp}})\rangle)$
 as a function of the rung coupling computed with truncation
 dimension $\mathbb{D} = 24$.
 The fidelity surface satisfies the symmetry property for exchanging
 $J_{\perp}$ and $J_{\perp}^{\prime}$.
 We find a pinch point at $(J_{\perp},J'_{\perp})= (0.383,0.383)$ in the parameter range
 $0.25 \leq J_{\perp} \leq 0.5$ with $J_\times = 0.2$ and $J_\parallel= 1$.
 The pinch point at $(J_{\perp},J'_{\perp})=(0.383,0.383)$ implies that a continuous phase
 transition occurs at $J_{\perp}^{\; c}\simeq 0.383$.
 Most importantly, only one pinch point in the fidelity surface
 is observed in the parameter range, which indicates
 that there are two characteristic phases.
 However, the CD phase was predicted to be intermediate between the rung-singlet and Haldane
 phases.\cite{SB04}
 In other words, the existence of the CD phase implies
 that there should be two pinch points in the parameter range.
 Consequently,
 the observed single pinch point on the fidelity surface
 indicates the non-existence of the CD phase in the parameter range.

\begin{figure}
\begin{center}
 \begin{overpic}[width=0.35\textwidth]{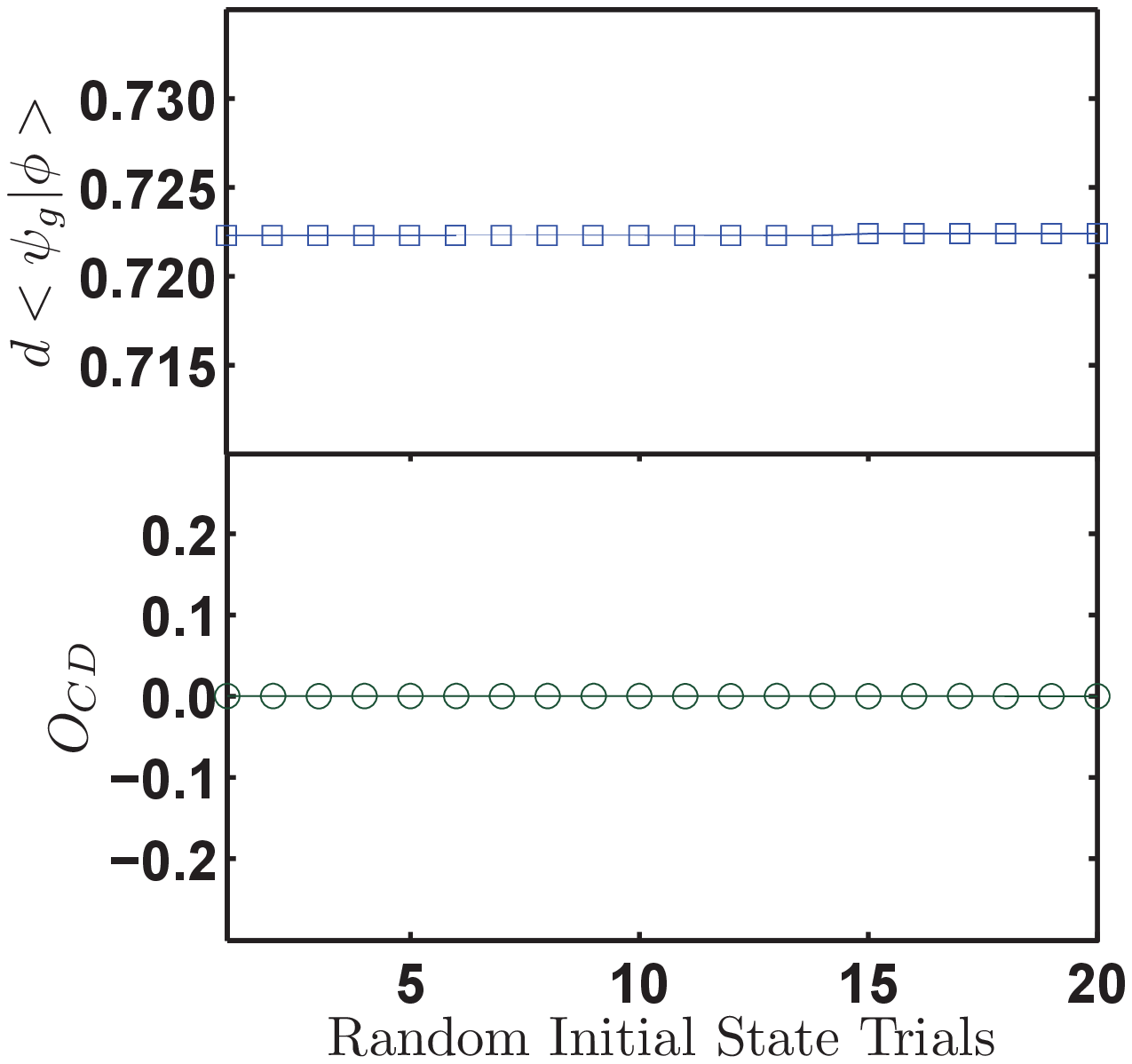}
 \put(-5,90){(a)}
  \end{overpic}
 \begin{overpic}[width=0.35\textwidth]{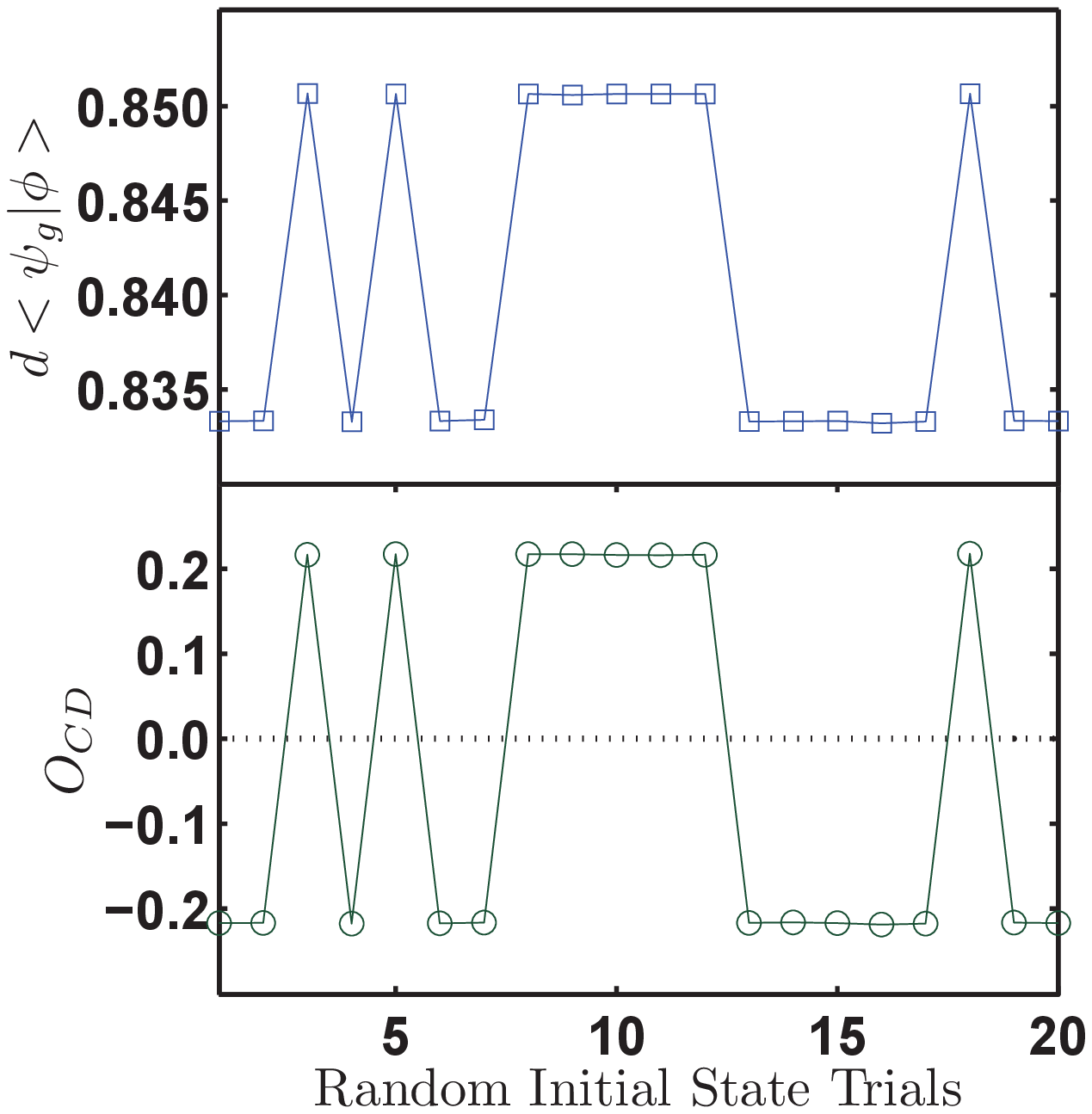}
 \put(-5,93){(b)}
  \end{overpic}
 \end{center}
\caption{(color online) Fidelity per lattice site $d$ between the
ground-state and a reference state.
The horizontal axis denotes the number $n$ of the random initial
state used to construct the ground-state.
Also show is the corresponding local order parameter $O_{CD}$.
(a) Antiferromagnetic parameter values $J_\parallel=1$,
$J_\perp=0.379$ and $J_\times=0.2$. In this case both the fidelity
per site and order parameter exhibit one distinct value indicative
of a non-degenerate ground-state.
(b) Ferromagnetic parameter values $J_\parallel=1$, $J_\perp=-1.6$
and $J_\times=-0.7$.
This point is known to be in the CD phase for ferromagnetic
couplings.
In the CD phase it is clearly observed that there exist two
degenerate ground-states corresponding to the two values of $d$ and
$O_{CD}$.} \label{fig3}
\end{figure}

\subsubsection{Detecting non-degenerate ground-state}

 One can make a further check on
 the existence of the CD phase,
 which has not been applied to ladder models before.
 According to the Landau theory for quantum phase transitions,
 if a phase can be characterized by a local order parameter, the
 phase is induced by a spontaneous symmetry breaking and, as a consequence, the system
 should have degenerate ground-states in the phase.
 If this theory is applicable for the CD phase,
 the CD phase should have degenerate ground-states.
Because the CD local order parameter \cite{HGCY06} is given by
\begin{equation}
 O_{CD}=\frac{1}{2}  \sum_{\alpha=1,2} \langle S_{\alpha, i}\cdot
 S_{\alpha, i+1}-S_{\alpha, i+1}\cdot S_{\alpha, i+2} \rangle  \,,
 \end{equation}
the CD phase should be responsible for two degenerate ground-states.
 It vanishes in the rung-singlet and Haldane phases.
This local order parameter should be nonzero in the CD phase and is
 thus a measure of the existence of the CD phase.

To determine the existence of the CD phase, then,
we need to detect degenerate ground-states.
The ground-state fidelity per lattice site
has been utilized in detecting degenerate ground-states for spontaneous symmetry breaking.
\cite{SHLC13,DCBZ14,WCB15}
To apply this scheme here, for a fixed system parameter,
 we need to calculate ground-states repeatedly with $n$ differently chosen initial
 states for the tensor network representation.
 Particularly, the system parameters are chosen to be
 $J_\perp=0.379$ with $J_\parallel=1$ for $J_\times=0.2$,
 because these values were suggested to be
 in the CD phase.\cite{HS10}
 For our numerical study, the initial states have been chosen randomly
 and used to calculate ground-states with the truncation dimension $\mathbb{D} = 12$.
 The reference state $|\phi\rangle$ has been also chosen randomly and has been kept to calculate the
 fidelity, which allows us to distinguish different ground-states
 with the same saturated energy in the numerical calculation.
 In Fig. \ref{fig3}(a) we plot the fidelity per lattice site $d$ for
 the number of initial trial states $n$.
  Twenty initial trial states are here presented to produce the same ground-state
 because there is only one observed value for the fidelity per lattice site $d$.
As a result, for this set of system parameters,
 the system does not undergo any spontaneous symmetry breaking
 and has a non-degenerate ground-state, which implies non-existence of the CD phase.
 The vanishing local order parameter $O_{CD}$ in Fig. \ref{fig3}(a) also manifests the non-existence of the CD phase.

 In contrast to the case of antiferromagnetic
 couplings, the CD phase is known to exist for
 ferromagnetic couplings ($J_\perp$ and $J_\times$ $< 0$).\cite{HS10,LSSLDZ12,LDW12}
 In order to compare with our results for antiferromagnetic couplings,
 we calculate degenerate ground-states for the
 parameter values $J_\parallel=1$, $J_\perp=-1.6$ and
 $J_\times=-0.7$.
 These values are known to be deep in the CD phase for ferromagnetic
 couplings.\cite{HS10,LSSLDZ12,LDW12}
 In Fig.~\ref{fig3}(b), we plot the fidelity per lattice site $d$ for
 the number of initial trial states $n$.
 In the CD phase it is clearly observed in the same way that there
 exist two degenerate ground-states corresponding to the two values of $d$.
 In fact, for a large enough number of random
 initial state trials, the probability $P(n)$ of each degenerate
 ground-state approaches $1/2$, i.e., $\lim_{n \rightarrow \infty} P(n)=1/2$.
 This gives an explicit detection of the two degenerate ground-states
 for the CD phase which originate from a $Z_2$ spontaneous symmetry
 breaking.
Moreover, as expected, the local order
 parameter $O_{CD}$ exhibits two distinct values that have the same
 amplitude and opposite sign, i.e., $O^{(1)}_{CD} = - O^{(2)}_{CD}$,
 where the local order parameters $O^{(1)}_{CD}$ and $O^{(2)}_{CD}$
 are calculated from the two degenerate ground-states $ |\psi_g^{(1)}
 \rangle$ and $ |\psi_g^{(2)} \rangle$, respectively.
 The relation between the two local order parameters
 manifests a $Z_2$ symmetry breaking for the two degenerate ground-states.
 Consequently, the sharp contrast between Fig.~\ref{fig3}(a)  and
 Fig.~\ref{fig3}(b) lends further weight to there being no CD phase
 for antiferromagnetic couplings.

\begin{figure}
\includegraphics [width=0.35\textwidth]{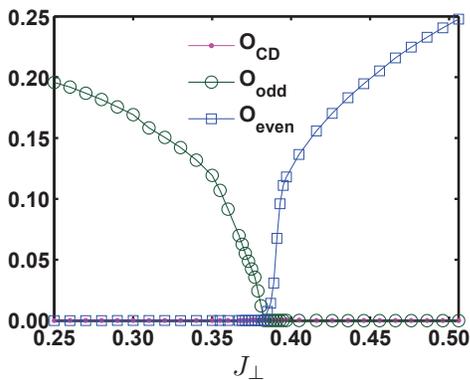}
 \caption{(color online) The string order parameters $O_\mathrm{odd}$
and $O_\mathrm{even}$ and the local order parameter $O_\mathrm{CD}$
as a function of $J_{\perp}$ for the cross-coupled spin ladder with
coupling $J_\times = 0.2$. Thin lines are guides to the eye. The
system clearly exhibits the Haldane phase for $J_{\perp} <  0.383$
and the rung singlet phase for $J_{\perp}>0.383$. The local order
parameter $O_{CD}$ is zero, indicating the absence of the CD phase.}
\label{fig4}
\end{figure}

\subsection{String order parameters}

 Order parameters for distinguishing
 between the rung-singlet and Haldane phases are
 nonlocal string order parameters.\cite{NR89}
 For two-leg spin ladders, we investigate the string order parameters
 $O_{\mathrm{odd}}$ and $O_{\mathrm{even}}$ defined by\cite{KFSS00,FLS01}
\begin{equation}
 O_{p}=-\lim_{|i-j| \to \infty }\langle S^z_{p,i}
 \exp \left( i\pi\sum_{l=i+1}^{j-1}S^z_{p,l} \right) S^z_{p,j}\rangle,
 \end{equation}
 where $p = {\mathrm{odd}}/{\mathrm{even}}$ with
 $S_{\mathrm{odd},i}=S^z_{1,i}+S^z_{2,i}$ and $S_{\mathrm{even},i}=S^z_{1,i}+S^z_{2,i+1}$.
 The order parameters $O_{\mathrm{odd}}$ and $O_{\mathrm{even}}$ are each non-zero in one phase
 and vanish in the other.
 More specifically, $O_{\mathrm{odd}}$ vanishes in the rung-singlet
 phase and $O_{\mathrm{even}}$ vanishes in the Haldane phase.
 Also,  $O_{CD}$ vanishes in the rung-singlet and Haldane phases.
 The behaviour of the three order parameters can
 thus also be used as a means of locating the quantum critical point and the range of each phase.

 In the previous subsection,
 we obtained the ground-states $|\psi(J_\perp)\rangle$
 as a function of $J_\perp$ for  $J_\parallel=1$
 and $J_\times=0.2$
 with truncation dimension $\mathbb{D} = 24$ to obtain the ground-state fidelity per lattice site in Fig. \ref{fig2}.
 In Fig. \ref{fig4}, we calculate and plot
 the order parameters from the ground-states in Fig. \ref{fig2}.
 Note that the local order parameter $O_{CD}$ is zero for the
 parameter range.
 Also, both the even/odd string order parameters $O_{even}/O_{odd}$
 vanish at $J_{\perp}^{\ c}\simeq 0.383$.
 The string order parameters clearly identify the Haldane phase for
 $J_{\perp} <  0.383$ and the rung singlet phase for $J_{\perp}>0.383$.
 As is detected from the pinch point on the
 fidelity surface in Fig.~\ref{fig2},
 the behavior of order parameters shows that a continuous phase transition
 occurs at $J_{\perp}^{c}\simeq0.383$.
 Consequently, the order parameters in Fig. \ref{fig4}
 show further that there is no evidence for the CD phase.

\section{Conclusion}

We have examined the phase diagram of the cross-coupled Heisenberg
spin ladder in Eq.(\ref{ham}) with antiferromagnetic couplings using
the recently developed tensor network states and ground-state
fidelity approach for quantum spin ladders.\cite{LSSLDZ12}
This method has been demonstrated to be particularly effective
 for identifying quantum phases and transition points.
We examined the particular coupling region $0.36 \le J_\perp \le
0.4$ for $J_\times=0.2$ for which there have been conflicting
results for the existence of the CD phase.
We identified a continuous phase transition at the value
$J_{\perp}^{\ c}\simeq0.383$ and clearly observed the Haldane and
rung-singlet phases.
However, we did not find any evidence for the existence of the CD
phase.
This conclusion was further supported by the use of a related
spontaneous symmetry breaking method, which we tested in the
vicinity of the critical point and for comparison in a region with
ferromagnetic couplings where the CD phase is known to exist.

A particular symmetry of the Hamiltonian in Eq. (\ref{ham}) poses a
strong constraint on the nature of the phase diagram for this model.
It was observed that $H(J_\parallel,J_\perp,J_\times)=H(J_\times,J_\perp,J_\parallel)$,\cite{ZKO98}
which is readily seen to be evident by inspection of the Hamiltonian.
Defining the variables $y_\perp=J_\perp/J_\parallel$ and $y_\times=J_\times/J_\parallel$ this symmetry implies
the ``duality" relation $H(1,y_\perp,y_\times)=y_\times H(1,y_\perp/y_\times,1/y_\times)$.
Thus the phase diagram for $y_\times \le 1$ maps to the values\cite{ZKO98,KLS08}
\begin{equation}
y_\perp \to y_\perp/y_\times, \qquad y_\times \to 1/y_\times .
\end{equation}
Consequently if the CD phase were to exist in the region $0.36 \le J_\perp \le 0.4$ for $J_\times=0.2$,
then there must also be phase boundaries in the corresponding region $1.8 \le J_\perp \le 2.0$ for $J_\times=5$.
Yet no further phase boundaries have been observed in this region, which is
consistent with the evidence for a single phase boundary, with no intermediate phase, between the rung-singlet and Haldane phases.
It thus seems plausable to rule out the existence of the CD phase in the antiferromagnetic region on symmetry arguments alone.

It appears then that the only contentious point remaining for the cross-coupled spin ladder with antiferromagnetic couplings
is identifying the order of the transition along the phase boundary.
It is clear that along this line a singlet-singlet gap vanishes between the gapped Haldane and rung-singlet phases.
The early analytic and numerical results were consistent with the transition being first order for both weak and
strong interchain couplings.\cite{ZKO98,KFSS00,FLS01}
However, later work suggested the transition was continuous for weak interchain coupling, becoming
first order for stronger coupling.\cite{W00,HGCY06}
It has also been concluded that the transition is first order along the whole line.\cite{KLS08}
Here we have seen that the transition is continuous at
$J_\times=0.2$, $J_\perp=0.383$.
This supports there being some crossover point(s) on the phase boundary separating lines of continuous and first order transitions.
Such a point is known to exist for the zig-zag ladder model, or
equivalently the $J_1$-$J_2$ spin chain, which exhibits a single
line of first order transitions for weak couplings crossing over to
second order transitions for strong couplings.\cite{KLS08}
The precise clarification of this issue for the cross-coupled spin ladder deserves further study.

{\it Acknowledgements.}
 XHC acknowledges support from the Fundamental Research Funds for the Central Universities
 (Project No.: 106112015CDJRC131215).
 MTB gratefully acknowledges support from Chongqing University and
 the 1000 Talents Program of China.
 This work is supported in part by the
 National Natural Science Foundation of China (Grant No. 11374379 and 11174375).


\end{document}